\documentclass[]{article}
\usepackage{graphicx}
\usepackage[latin1]{inputenc}
\usepackage{amsmath}
\usepackage{comment}
\usepackage{xcolor}
\usepackage[urlcolor=blue]{hyperref}      % links to citations
\hypersetup{
    colorlinks = true,                    % text and not border
    citecolor = {blue},
    linkcolor = {purple},
           }
\title{\bf Incentives for accelerating the production
of Covid-19 vaccines in the presence of adjustment
costs\footnote{This research has 
been supported by  the European 
Union's Horizon 2020 research and
innovation program, PERISCOPE: 
Pan European Response to the Impacts 
of Covid-19 and future Pandemics and
Epidemics, under 
grant agreement No. 101016233, H2020-SC1-PHE CORONAVIRUS-2020-2-RTD.
The authors thank Roser Valent\'i\ for 
1

valuable comments.
} }

\author{Claudius Gros\footnote{Professor, Institute for Theoretical Physics, Goethe University Frankfurt, Germany}, 
Daniel Gros\footnote{Distinguished Fellow, CEPS (Centre for European Policy Studies), 
Brussels, Belgium} 
}

\begin{document}
%%%%%%%%%%%%%%%%%%%%%%%%%%%%%%%%%%%%%%%%%%%%%%%%%%%%%%%%%

\maketitle
\begin{abstract}
Delays in the availability of vaccines are costly as 
the pandemic continues. However, in the presence of 
adjustment costs firms have an incentive to increase 
production capacity only gradually. The existing 
contracts specify only a fixed quantity to be supplied
over a certain period and thus provide no incentive for an
accelerated buildup in capacity. A high price does not 
change this. 
The optimal contract would specify a decreasing price schedule
over time which can replicate the social optimum.
\end{abstract}

%%%%%%%%%%%%%%%%%%%%%%%%%%%%%%%%%%%%%%%%%%%%%%%%%%%%%%%%%
%%%%%%%%%%%%%%%%%%%%%%%%%%%%%%%%%%%%%%%%%%%%%%%%%%%%%%%%%

%===========================
%\thispagestyle{empty}    % manually for the first page
%\newpage
\section{Introduction}
%===========================

As governments grapple with the second wave 
in Europe, they also start mass scale vaccination 
campaigns, hoping to achieve herd immunity, which 
is the point at which a high enough percentage of 
the population has been vaccinated so that the 
virus will abate. However, vaccination takes time 
because increasing capacity is subject to adjustment 
costs. The supply of vaccines is thus limited in the 
short run \cite{Vellodivacpolicycovideconomics}.

Adjustment costs lead firms to 
increase capacity only gradually, which might not 
be optimal from a social point of view. The problem 
for public authorities is then to find a way to accelerate
the increase in production capacity. The vaccine 
supply contracts were mostly concluded before the 
vaccines had even been fully developed, let alone
approved for general use. It was thus impossible to 
impose a tight deadline for delivery. The Advance 
Purchase Agreements of the EU, two of which have been 
published \cite{AstraZenecacontractEU,CurvaccontractEU},
thus specify only an overall price and tentative 
delivery schedules in terms of quarters, not months 
or weeks. When even these tentative schedules start 
to slip (as is the case now), the EU has little 
leverage to induce companies to make efforts to 
accelerate delivery. $\footnote{The Heading of the AstraZeneca contract is  "ADVANCE PURCHASE AGREEMENT (APA) for the production PURCHASE AND SUPPLY OF A COVID VACCINE IN THE EUROPEAN UNION. A large part of the APA is devoted to the question that the vaccine be produced in the EU (and not elsewhere).  To ensure production within the EU seems to have been the focus of the Commission at the time.}$ We analyze the consequences of 
this type of contracts for the supply schedule of a 
vaccine and how the resulting incentives for back-loading 
supplies can be mitigated.

Limitations: A substantial part of the literature on vaccine policy 
focuses on how and whom to vaccinate, usually taking 
it as granted that the supply of vaccines is not a 
constraint \cite{Vellodivacpolicycovideconomics,grauer2020strategic,anderson2020challenges,wang2016statistical}.
We do not consider this issue as we 
concentrate on the case of Covid vaccines, for which 
mass production had to start immediately after 
test trials were successful. Another, issue we do not consider is 
vaccine hesitancy \cite{dube2013vaccine}, 
leading to doubts about whether herd immunity will be achieved 
if a certain proportion of the population refuses the 
vaccine. The immediate problem facing policy makers is 
the opposite, at least initially. The demand for 
vaccination far outstrips supply. Moreover, even if full 
herd immunity could not be reached, there is still a 
considerable benefit from every person vaccinated, which 
reduces potential hospitalization costs correspondingly,
allowing governments in the aggregate to end lockdown 
measures earlier \cite{ahuja2020preparing,cutler2020covid}. 

Here we do not consider the issue of uncertain efficacy 
of vaccines and the related problem of 
ordering portfolios of potential candidates 
\cite{ahuja2020preparing}, as done by most
major countries. Neither do we consider uncertainties
in production costs \cite{abel1997exact}.

Our analysis concentrates on the problem of ramping 
up production once the efficacy of a vaccine has been 
established \cite{khamsi2020if}. The importance of this 
issue for the global economy has been laid out in
\cite{cutler2020covid}. 

We start by analyzing the case in which firms 
producing the vaccines optimize the production 
time path with the aim of minimizing costs under
the constraint of a fixed quantity to be produced
within a given period of time (and given a fixed price).  
We also consider the problem of building up production 
over time from the perspective of a social planner 
and show that it is equivalent to a pricing scheme 
that is linear in time. In this case an initially 
high starting price declines over time. 
The resulting optimal pricing scheme aligns the 
interests of the producer with that of the society 
as a whole.

%===========================
\section{Adjustment costs for ramping up vaccine production}
%===========================

The problem that the producers of a new product, 
like the Covid vaccines, face involves one key element, 
namely adjustment costs. It is not possible to ramp
up production instantaneously. Standard economic analysis 
takes this into account by positing that there is a 
cost to increasing capacity and that this cost is convex, 
i.e.\ the costs of increasing capacity are only
small when the buildup is slow
\cite{hamermesh1996adjustment,cooper2006nature} .
The implication is straightforward: it will be
optimal to smooth production over time.

Consider a contract in which a certain quantity
$Z_T$ is to be delivered over a period $T$ (say
one year), at a constant price $p_0$ per unit. In 
this case the exact timing of the delivery, close 
to the start or to the end of the delivery period, 
does not matter and it will be optimal for the 
producer to minimize costs by increasing capacity only
gradually over time. 

Denoting the instantaneous production capacity (the
number of vaccines produced per unit time, say daily) 
with $z_t$, the adjustment costs will be a function
$f(\dot{z}_t)$ of $\dot{z}_t$, which quantifies the
speed at which production is ramped up. Overall adjustment costs 
are then determined by the integral of $f(\dot{z}_t)$ 
over the delivery period, subject to the constraint
that a total of $Z_T=\int_0^T z_t dt$ units are produced.

In most applications \cite{hamermesh1996adjustment},
the convexity of the costs of adjustment is assumed 
to be quadratic (which would also be the result 
of a second order approximation). With quadratic 
adjustment costs, as considered here, the marginal 
cost of adjusting becomes linear, allowing for explicit 
solutions.

%---------------------------
\subsection{The time path for production
under fixed price contracts\label{sect_constant_price}}
%---------------------------

We start by analyzing the production path 
resulting from the type of contract that has been 
used so far, namely a fixed price against delivery 
of a certain quantity over a time period specified 
in advance. For example, the Advanced Purchase 
Agreement of the European Union with Curevac 
specifies the delivery of certain amounts of doses 
for the year 2021 \cite{CurvaccontractEU}, with 
only \emph{tentative} delivery schedules by quarter. 
This implies that the firm can distribute the 
supply schedule over the entire year, which is 
nearly an eternity in terms of a pandemic costing
several percentage points of GDP at each instant,
threatening at the same time the lives of thousands 
every day. The EU contract with AstraZeneca \cite{AstraZenecacontractEU} specifies only \emph{reasonable best efforts} and the one with Sanofi \cite{SanoficontractEU} contains a similar formula.

We thus focus on the inter-temporal 
problem of increasing production capacity over time 
within the overall time frame given by the contract, 
which could be thought of representing one year. 
Given this time frame (and interest rates around zero), 
we neglect time discounting.

Formally we consider a firm which has been contracted 
to supply a certain amount $Z_T$ of doses over a given 
period $T$. The marginal cost of each dose is denoted by c
and is assumed 
to be constant once the capacity has been created.

Capacity means in this case not just the physical 
factory, which might have to satisfy specific 
requirements, but also the schooling of personnel, 
etc. We assume that the initial capacity is low, 
possibly equal to zero, but definitely not large 
enough to satisfy the entire order within 
$t\in[0,T]$. This implies that the firm must ramp 
up capacity during the contract period $[0,T]$.

The problem for the firm is then to maximize 
revenues minus adjustment costs, 
subject to the overall production constraint:
\begin{equation}
p_0 \int_0^T z_t dt - a_z\int_0^T (\dot z_t)^2 dt
- \lambda_Z\left[\int_0^T z_t dt-Z_T\right] - c \int_0^T  z_t dt\,,
\label{adjustment_costs_p const}
\end{equation}
where $p_0$ denotes the price per vaccine,
$a_z$ encodes the size of the adjustment 
cost, and where $\lambda_Z$ is the Lagrange 
multiplier enforcing the constraint that the 
total production over the period $[0,T]$ 
is $Z_T = \int_0^T z_tdt$. 

We use adjustment costs in absolute, not relative terms.  This means that these costs do not depend on 
the level of capacity already reached. Substantial
effort has been devoted in the literature to the 
study of adjustment costs modelling them in terms of a proportional 
increase in capacity \cite{hamermesh1996adjustment}, 
i.e.\ adjustment costs that are a function of 
$\dot{z}_t/z_t$. However, this would lead to 
conceptual difficulties when starting the production 
of a new product (i.e.\ when $z_0 = 0$), 
the case of Covid-19 vaccines. 

This formulation (\ref{adjustment_costs_p const}) 
of transactions costs is assumed to reflect roughly 
real world technical constraints. That `overnight 
factory constructions' are not possible is translated
in the framework of equation (\ref{adjustment_costs_p const}) into adjustment 
costs that diverge to infinity. Standard variational calculus 
\cite{troutman2012variational,gros2015complex}
establishes that the stationary solution 
to (\ref{adjustment_costs_p const}) satisfies
\begin{equation}
2a_z\ddot{z}_t = \lambda -p_0 - c,
\qquad\quad
z_t = z_0 + \gamma t +\frac{\lambda-p_0 - c}{4a_z}t^2 \,,
\label{z_t_variational price constant}
\end{equation}
where $z_0$ is the initial production capacity and 
$\gamma$ the speed at which production capacity 
increases initially. 

Only the difference between price and marginal costs enters the condition \ref{z_t_variational price constant}.  In the remainder we thus assume that c is equal to zero. 
This assumption is made only for computational convenience.
Any constant marginal cost would only add a fixed amount to the overall
costs of the firm because the total quantity to be produces is fixed.  One could thus think of the price as 
representing the difference between the unit price contracted 
and any marginal cost of production.

Production capacity remaining at the end of the production period
is worthless.  This assumption can be 
However, our results would not change even if production capacity at the end of the 
contractual period were to have a value because this value
would just add a constant term to the firms revenues and 
would thus not affect the time path for the build-up of
capacity, which is our main object of analysis.

The problem that the firm faces can be reduced 
to minimizing the total cost of adjustment 
over the delivery period $T$, as total revenues 
are fixed, being equal to the price times 
the quantity delivered. The production schedule 
that minimizes the adjustment cost is to increase 
capacity accordingly to (\ref{z_t_variational price constant}).  

A constant rate of increase in production would not be 
optimal, on general grounds, because an increase in 
capacity implemented today yields higher production 
over the remainder of the delivery period and is thus 
more valuable that an increase in capacity just 
before the end of the delivery period. The speed at 
which capacity increases should thus decline over time.
This intuition is born out by
(\ref{z_t_variational price constant}).

To be concrete, we parameterize the
solution to (\ref{z_t_variational price constant}) 
with
\begin{equation}
z_t = z_0 + \gamma t + \delta t^2,
\qquad\quad
\delta = \frac{\lambda-p_0}{4a_z}\,.
\label{z_t_contant_price}
\end{equation}
The production condition $Z_T = \int_0^T z_tdt$ 
implies then
\begin{equation}
Z_T = z_0 T + \frac{\gamma}{2} T^2 + \frac{\delta}{3} T^3,
\qquad\quad
\gamma = \frac{2\Delta Z}{T}-\frac{2\delta T}{3}\,.
\label{Z_T_shadow}
\end{equation}
where $\Delta Z$ denotes the difference between 
the average capacity needed to fulfill the order 
and the initial one, $ \Delta Z = Z_T/T-z_0$. It is
assumed here that $\Delta Z>0$, namely that the capacity
needs to be increased. In the opposite case, when $Z_T<Tz_0$,
the company would have to shut down part of the existing
production capacity - which is not the case for Covid vaccines.

\begin{figure}[t!]
\centerline{
\includegraphics[width=0.9\textwidth]{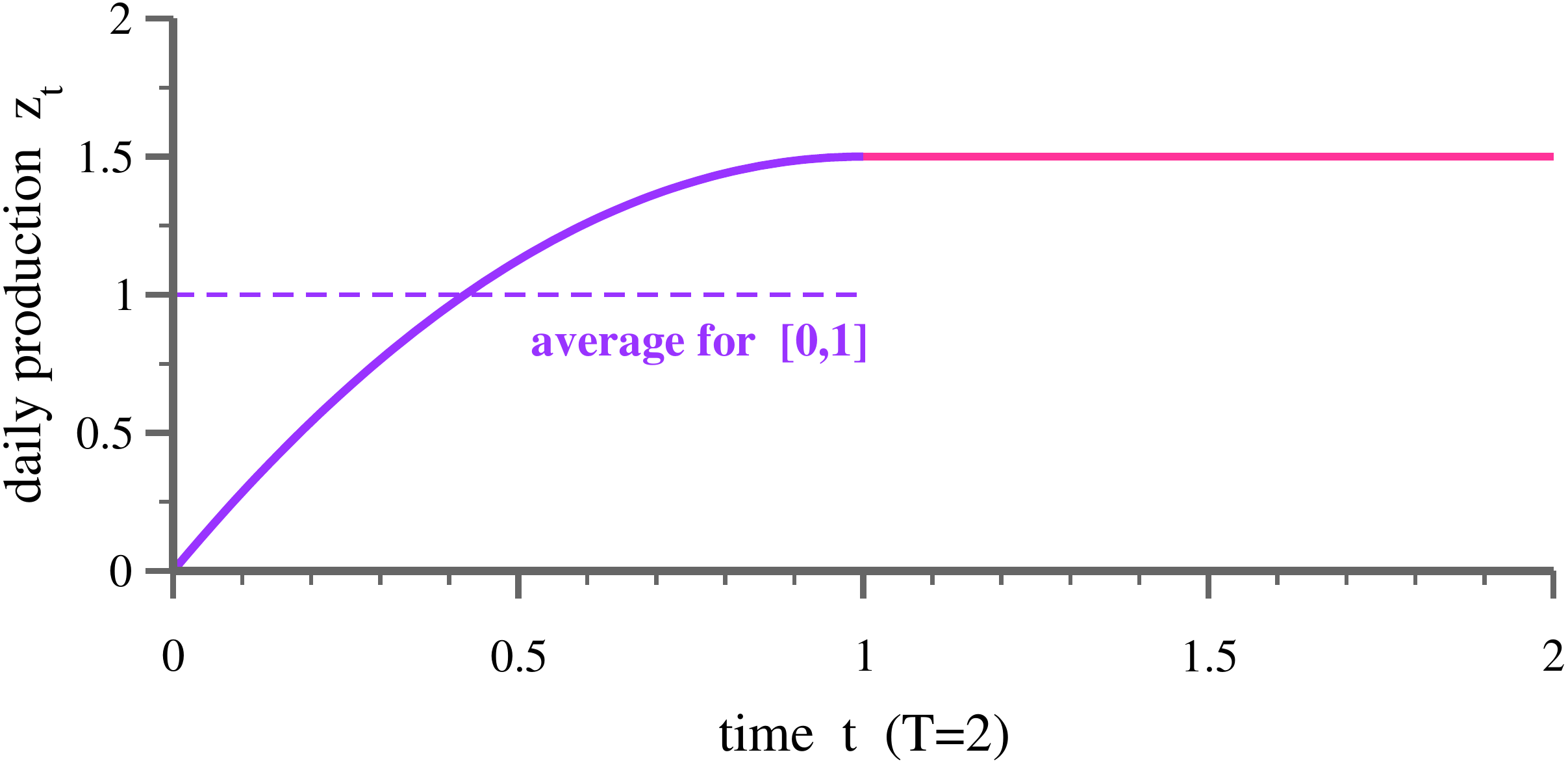}
           }
\caption{The time evolution of the production capacity
$z_t$, taken here to represent the daily output.
In the first period, $t\in[0,1]$, an average
of $\langle z_t\rangle=1$ is to be attained
(in this example). Minimization of the adjustment 
costs leads to (\ref{z_t_min}), which is shown. 
Note that $z_{t=1}$ overshoots the average by
50\%. Production per time remains constant
in the second period $t\in[1,2]$. During
the overall delivery time $T=2$ the production 
capacity is on the average $(1+3/2)/2=1.25$.
}
\label{fig_adjustmentCosts_z_t}
\end{figure}

The overall production constraint (\ref{Z_T_shadow}) 
can be satisfied by any linear combination of 
$\gamma$ and $\delta $. These two parameters are 
determined by maximizing total profit. Given that 
the first term in (\ref{adjustment_costs_p const}) 
is constant, $p_0Z_T$, one just has to minimize the
adjustment costs:
\begin{equation}
E_{\rm adj} =
a_z\int_0^T (\dot{z}_t)^2 dt = 
a_z \left[ \gamma^2 T + 2\gamma \delta T^2
+ \frac{4\delta^2 T^3}{3} \right] \,,
\label{E_adjustments min}
\end{equation}
where $\dot z_t = \gamma + 2\delta t$ has
been used.
The relation (\ref{Z_T_shadow}) entails
that $\partial\gamma/\partial\delta = -2T/3$,
which leads to
\begin{eqnarray} 
\label{d_E_adj_delta}
\frac{dE_{\rm adj}}{d\delta} &=& 
\frac{\partial E_{\rm adj}}{\partial\delta}
+\frac{\partial E_{\rm adj}}{\partial \gamma}
\frac{\partial \gamma}{\partial \delta}
= \frac{\partial E_{\rm adj}}{\partial\delta} -
\frac{2T}{3}\frac{\partial E_{\rm adj}}{\partial\gamma}
\\[0.5ex]
&=&   a_z\left[2\gamma T^2+\frac{8\delta T^3}{3}
-\frac{2T}{3}\big( 2\gamma T+2\delta T^2 \big)\right]
\\[0.5ex]
&=&   a_z\left[
\frac{2\gamma T^2}{3}+\frac{4\delta T^3}{3}
\right]=0\,.
\nonumber
\end{eqnarray}
The first order condition for cost minimization 
over the choice of $\gamma$ and $\delta$ leads 
therefore to following simple relationships:
\begin{equation}
\gamma = -2 \delta T,
\qquad\quad
\gamma = 3 \frac{\Delta Z}{T},
\qquad\quad
\delta = - \frac{3 \Delta Z}{2T^2}  \,.
\label{delta_gamma}
\end{equation}
where the last two relations follow from
(\ref{Z_T_shadow}). The time path $z_t$
for the production capacity is then
\begin{equation}
z_t =  z_0  + \frac{3\Delta Z}{T}\left[
t - \frac{t^2}{2T}\right],
\qquad\quad
\dot z_t = \frac{3\Delta Z}{T} 
\left[1 - \frac{t}{T}  \right]\,,
\label{z_t_min}
\end{equation}
as illustrated in Figure~\ref{fig_adjustmentCosts_z_t}.
This implies that production capacity follows an inverted
parabola, with the highest level reached just before the end
of the delivery period. The \emph{increase} in capacity starts strongly,
but declines over time
tending towards zero at the end of the delivery,
$\lim_{t\to T}\dot{z_t}\to0$. The result (\ref{z_t_min})
also implies that $\dot z_t$ is proportional
to the missing average production capacity $\Delta Z$,
scaling inversely with the production period
$T$. At the end of the contract period (t=T), 
the production capacity will be equal to 1.5 times 
the one which is needed on average ($Z_T/T$ when
$z_0=0$).

Note that the cost minimizing production path $z_t$ does 
not depend on the overall value of the order since the price
$p_0$ does not influence the parameters of the differential 
equation (\ref{z_t_min}). The reason is that $p_0$ enters 
the Lagrange multiplier of the equation of motion 
(\ref{Z_T_shadow}) only through the difference 
$\lambda-p_0$. 

The key corollary from the above considerations, regarding
the effects of adjustment costs, is then: 
\begin{quote}
The level of the price does not influence the speed at 
which production is increased -- when the price is 
constant. 
\end{quote}
Higher prices allow the producer to obtain larger 
profits, however without providing incentives to 
accelerate the buildup of the capacity.
We have not considered explicitly the cost of developing the vaccine, which would add a constant term to the costs for the firm.
But this constant term would also not have any influence on the 
speed at which production is increased since it represents just
a sunk cost when the firm starts to ramp up production. 

%=====================================
%=====================================

%----------------------------
\subsection{Adjustment costs scaling}
%----------------------------

We have established so far that a constant price 
does not affect the time path of production,
but the firm will accept the contract only if total revenues,
$Z_T p_0 $
compensate all costs costs, $E_{\rm adj}$, 
which can be obtained by substituting 
(\ref{delta_gamma}) into
(\ref{E_adjustments min}).

\begin{equation}
E_{\rm adj} = 3 a_z \frac{\Delta Z^2}{T}
= 3 a_z \frac{(Z_T-z_0T)^2}{T^3}\,,
\label{E_adj_final}
\end{equation}
Total adjustment costs are linear in the adjustment 
cost parameter, $a_z $. 
For $z_0=0$, ceteris paribus, 
they fall with the cube of the time 
the firm has to fulfill the entire order. 
A positive level of initial production $z_0 > 0$ helps to
reduce adjustment costs and this effect increases 
with the length of the time period given for delivery, 
$T$. These scaling relations hold for
fixed overall production $Z_T$. 

The equation \ref{E_adj_final} represents also the minimum total expenditure 
public authorities would have to sustain 
in a market in which firms compete for vaccine orders.  (Development costs were separately financed.) 
The EU has concluded contacts with 6 suppliers of vaccines
which appeared to have a realistic chance of success in 2020.
There was thus some, but also certainly not perfect, 
competition across different vaccine producers.  

We do not take a stance on how competitive the vaccine market
is and whether the prices actually paid reflected mainly costs. However, one can still recover from the expression for the total adjustment costs a lower bound; i.e. the unit price needed to induce a firm to accept the contract.\footnote{We recall that the price was defined as the difference to the marginal costs of production.  The shadow prices discussed here should thus be increased by marginal costs. The same argument can also be applied if the capacity build up to satisfy this one-time order has a scrap value.  This value will not affect the time path of production, but it will affect the price at which the firm can break even.} This is given for $z_0=0$ by:
\begin{equation}
\frac{E_{\rm adj}}{Z_T}  = 3 a_z \frac{Z_T}{T^3}
\label{E_adj_final unit}
\end{equation}

A first corollary of (\ref{E_adj_final unit}) is that, for a given amount ordered, $Z_T$, the unit
cost, and thus the shadow reservation price, increases with the cube  Halving T requires an 8-fold higher price to compensate for rapidly increasing adjustment costs.
if a higher number of doses has to be delivered by the same time.
The price would still have to increase four times price when halving both the delivery time $T$ and the total amount 
requested $Z_T$, thus keeping average delivery intensity constant. 
Moreover, given T, a larger order requires a higher unit price.
The underlying reason is that capacity has to be increased faster
when starting from zero.

We note that the overall delivery time $T$ is not the same as 
the average delivery delay $t_{\rm deliver}$,
over the life-time of the product, which is
given by 
\begin{equation}
t_{\rm deliver} = 
\frac{1}{Z_T} \int_0^T z_t\, t\, dt =
\frac{1}{Z_T/T} \left[
\frac{z_0}{2} + \frac{5\Delta Z}{8}  \right] T\,.
\label{T_deliver}
\end{equation}
With $\Delta Z=Z_T/T-z_0$ this result
implies that for $z_0=0$, i.e.\ when 
the initial capacity is zero and $\Delta Z=Z_T/T$, 
the mean delivery delay will be $5T/8$. Instead, 
when $\Delta Z = 0$, viz when production is 
constant, the average delay would be $T/2$. 
The later case holds when the initial production 
capacity is sufficient to fulfill the entire 
order over time, which is however not the 
case for Covid-19 vaccines.

%----------------------------
\subsection{Social costs of delay vs.\ adjustment costs}
%----------------------------

For society the value of a dose depends 
importantly on the time it is delivered. 
Early delivery helps to avoid infections 
and allows for an earlier lifting of 
lockdowns. This implies that any delivery 
which does not occur today imposes an 
opportunity cost for society which is 
due to the economic loss of prolonged 
lockdowns and more infections. Each 
early dose thus delivers a flow of 
benefits in terms of avoided costs which is 
proportional to the time it arrives. This 
benefit does not materialise only when 
herd immunity has been reached. Every 
person vaccinated will reduce the
potential medical costs from an infection. 
We thus assume that the opportunity 
cost to society of the 'no vaccination' status quo
is $k$ per unit of time, where $k$ parametrizes the per unit 
costs of lost output, due to a continuing lockdown.  
Assuming that the total amount ordered, $Z_T$ is sufficient 
to stop the pandemic, the cost is
reduced pro rata the part of the population which
has been vaccinated, $\frac{z_t}{Z_T}$

The aim for society should be to minimize 
the sum of social opportunity costs of 
delay and the adjustment costs in ramping 
up vaccine production. For the case of a 
contract with a fixed price, the total 
adjustment cost were calculated in 
(\ref{E_adj_final}). This can be combined
with the time path of capacity (which 
is proportional to the number of people 
vaccinated each period) to yield:

The economic costs are proportional to the
relative number of people yet not vaccinated, 
\begin{eqnarray}
\nonumber
k \int_0^T \left(1- \int_0^t \frac{z_{t'}}{Z_T}dt'\right)\,dt &=&
k \int_0^T \left(1-\frac{3}{2}\frac{t^2}{T^2}
+\frac{1}{2}\frac{t^3}{T^3}\right)\,dt
\\[0.5ex]    &=&
\frac{5}{8}kT\,,
\label{economic_costs_alone}
\end{eqnarray}
where (\ref{z_t_min}) has been used, together 
with $z_0=0$. Over the same period the costs 
would be $kT$ if nobody would be ever vacinated,
which implies that ordering vaccines at a fixed 
pice leads to a reduction of opportunity costs
of $3/8$ already within the delivery period, together
with 100\% reduction afterwards.

Overall social costs $E_{\rm social}$ 
are the economic costs (\ref{economic_costs_alone}) together 
with the adjustment costs (\ref{E_adj_final}),
\begin{equation}
E_{\rm social}
= \frac{5}{8}kT + 3 a_z \frac{Z_T^2}{T^3}\,,
\label{Social_costs}
\end{equation}
where (\ref{E_adj_final}) and (\ref{T_deliver}) 
have been used for the case $z_0=0$. The 
delivery period $T$ minimizing the sum of 
social and adjustment costs is hence
given by
\begin{equation}
T_{\rm opt} = \left[\frac{72Z_T^2}{5}\frac{a_z}{k}\right]^{1/4}\,.
\label{T_deliversocialoptimal}
\end{equation}

Using the solution for the optimal time (\ref{T_deliversocialoptimal})
 in the expression for the 
unit adjustment costs 
(\ref{E_adj_final unit}) yields a solution for the price at 
which the firm would be willing to supply:
\begin{equation}
\frac{E_{\rm adj}}{Z_T} = 3 a_z \frac{Z_T}{T^3} = 
3a_z Z_T^2\left[\frac{5k}{72a_z Z_T^2}\right]^{3/4}=
3 a_{z}^\frac{1}{4}\left[ Z_T\right]^{-1/2} \left[\frac{5k}{72}\right]^{3/4}\,.
\label{E_unit optimal}
\end{equation}

This implies that the unit cost for the firms
increases steeply with 
the total amount ordered (exponent $1/4$),
and almost linearly (exponent 3/4) with the
social cost factor k. The unit cost at which
firms would supply (in the context of a fixed
price contract) thus does not depend only 
on the \emph{ratio} of these two parameters.  
We will return to the 
relative size of adjustment and social costs 
in Sect.~\ref{sect_orders_of_magnitude}.

%---------------------------
\subsection{Societal gains from vaccination and asymmetric information}
%---------------------------

The unit costs in (\ref{E_unit optimal}) reflect only the
minimum needed for the firm to break even.
It does not reflect the amount the authorities 
should be willing to pay, which should be function
of the total societal costs avoided. Without a 
vaccine the pandemic would continue until
most people have been affected and natural herd 
immunity has been reached. With continuing NPIs, 
this would not happen in an uncontrolled exponential 
growth, but gradually over time. For example, in 
the US, at least 8 percent of the population has 
been infected (in the sense that 8 percent had a 
positive test result)  over one year.  This implies 
that it would take several years under similar 
conditions for natural herd immunity to be reached.
The gain to society from the availability of a 
vaccine should thus be measured against a baseline 
of the costs continuing not until T, but until until 
natural herd immunity is reached, $T_{nhi}$. The total 
gain to society would thus be equal to k times the 
period of zero cost after full vaccination (which is 
equal to $T_{nhi}-T$) plus the gains reaped during 
the vaccination period.

The latter can be calculated assuming that the 
delivery time, $T$, is set by the authorities 
according to equation (\ref{T_deliversocialoptimal})
the relationship. Substituting this back into 
the expression for the overall social cost 
(\ref{Social_costs}) yields an expression for 
the `minimum social cost', i.e., the social 
costs that remains even if the time has been set 
so as to minimize social costs:
\begin{equation}
E_{\rm min\ social\ costs}
= T \Bigg[\frac{5}{8}k + 3 a_z \frac{Z_T^2}{T^4}\Bigg] =
T k \bigg[\frac{5}{8} + \frac{15}{72} \bigg] = 
\frac{5Tk}{6}\,.
\label{Social_costs_opt}
\end{equation}
This implies that (with a fixed price contract), only 
one sixth of the cost of the pandemic that arises during 
the period of increasing production to vaccinate the
population can be avoided.
\begin{equation}
{\rm SocialBenefit}
= k \left[T_{nhi} - T\right] + \frac{Tk}{6} =
T k \left[\frac{T_{nhi}}{T} - \frac{5}{6}\right]\,,
\label{Social_benefits}
\end{equation}
The ratio of social benefits to overall expenditure
needed to obtain the necessary amount of vaccines
in the optimal period can now be calculated using
(\ref{z_t_min}) together with 
(\ref{T_deliversocialoptimal}) as:
\begin{equation}
\frac{\rm SocialBenefit}{\rm Cost\ of\ order} 
= kT \frac{\frac{T_{nhi}}{T} - \frac{5}{6}}{3 a_z \frac{Z_T^2}{T^3}}
=
\frac{24}{5}\left[\frac{T_{nhi}}{T} - \frac{5}{6}\right]\,.
\label{Social_benefits_cost}
\end{equation}
This implies that the social benefits should be a multiple
of the cost of placing the order (provided, of course, 
that the time needed to reach natural herd immunity would be
longer than the time needed to reach this goal through
vaccination). For example, social benefits would be 
twenty times larger than the cost if vaccination would
reduce the time of the pandemic by one fifth.  

This potentially very large relative difference between 
social benefits and the private shadow price would 
become important if there is asymmetric information. 
The authorities would be willing to pay a much higher 
price than would be necessary to induce companies to 
supply the vaccine. 

We have so far considered only contracts which 
specify a fixed price. The optimal contract time 
calculated in (\ref{T_deliversocialoptimal}) above
constitutes a second best, because it is subject to 
this constraint. We now turn to the optimal contract 
design when this constraint is lifted.

%===========================
\section{Optimal time-varying pricing}
%===========================

Using the expression for the opportunity costs 
of delay introduced above in (\ref{economic_costs_alone}), 
the general social planner problem, which is 
not constrained by a fix price contract, is to minimize 
the sum of the costs of an ongoing lockdown and 
the adjustment costs that are necessary to 
accelerate production. The end point, $T$, 
represents the point in time when herd immunity
has been reached, i.e. when a high per percentage of the 
entire population has been vaccinated.
At this point economy would be fully back to normal 
and the costs parametrized by k no longer arise.
It is usually assumed that for Covid-19 herd immunity requires that about 70 percent are vaccinated.  We approximate this by normalising $Z_T=1$.

%======================
\begin{comment}
%
\begin{equation}
W_{\rm social} = k\int_0^T 
\left(1- Z_t\right)dt +
a_z\int_0^T (\dot{z}_t)^2 dt - 
\lambda_Z\left[\int_0^T z_t dt-1\right]\,,
\label{Social_planner}
\end{equation}
%
where $Z_t=\int_0^t z_{t'}dt'$ is the number
of vaccines produced hitherto, which is assumed
to correspond to the proportion of people already 
vaccinated at that point.
\end{comment}
%======================

Denoting total social costs by $W_{\rm social}$, 
the social planner takes in account the opportunity 
costs of gradual delivery, which are proportional to the 
time one waits for the delivery of the vaccine:
\begin{equation}
W_{\rm social} = 
k \int_{0}^{T} t z_t dt +
a_z\int_0^T (\dot{z}_t)^2 dt - 
\lambda\left[\int_0^T z_t dt-1\right]\,.
\label{Social_cost}
\end{equation}
It can now been shown that the problem for 
the social planner can be made isomorphic 
to that of the firm. The key variable for 
the firm is the price, or revenue per unit 
produced. In a fixed price contract this 
price does not vary with the time the 
vaccine is delivered. This can be changed 
if the authorities offer a time varying price 
for example one which declines from a certain 
initial level. With a price $p_t$ that is 
variable over time, the total revenues of the 
firm are given by:
\begin{equation}
Revenue = \int_{0}^{T} p_t z_t dt\,.
\label{Revenues}
\end{equation}
For both, the social planer and the firm, the problem
has to be solved taking into account adjustment
costs. The social optimum of (\ref{Social_cost}) can 
be reached if the price path facing the firm coincides 
with the minimization of the pandemic costs, i.e.\ if
\begin{equation}
p_t = p_0 - kt\,,
\label{pricing_schedule}
\end{equation}
where $p_0$ now denotes the `base price', 
which diminishes linearly over time.
The problem facing the firm then becomes 
to maximize total revenues minus 
adjustment costs:
\begin{equation}
\int_0^T \big(p_0-kt\big) z_t dt - a_z\int_0^T (\dot z_t)^2 dt
- \lambda\left[\int_0^T z_t dt-1\right]\,,
\label{adjustment_costs_premium}
\end{equation}
which can be rewritten as:
\begin{equation}
- k \int_0^T t z_t dt - a_z\int_0^T (\dot z_t)^2 dt
- (\lambda-p_0)\int_0^T z_t dt + \lambda\,.
\label{adjustment_costs_firm}
\end{equation}
Comparing equations (\ref{Social_cost}) and
(\ref{adjustment_costs_firm}) shows that
they lead to the same solution, vzi to
the time path for $z_t$. Note that the sign of
the Lagrange parameter $\lambda$ changes, but
this is irrelevant. The firm \emph{maximizes} the
difference between revenues and adjustment
cost, with unit revenues declining linearly
over time. Society \emph{minimizes} total costs,
which comprise the same adjustment costs,
but taking also into account that the costs
of delayed delivery are linear in time. 
With the pricing schedule (\ref{pricing_schedule}), 
equations (\ref{Social_cost}) and 
(\ref{adjustment_costs_premium}) represent 
hence the same problem, except for the constant 
term $p_0$, which implies that they have the 
same solution $z_t$. The size of the initial 
price, $p_0$, has no consequences for the 
decision of the firm regarding how
quickly to increased capacity.

The implication is that the pricing schedule
(\ref{pricing_schedule}) can induce firms 
to adopt the speed of increase in production 
capacity which is also optimal from a social 
point of view. There is thus a way to align 
private and public interests by specifying 
a pricing schedule which mimics the social 
cost of a continuing pandemic. 
The base price $p_0$ determines, as before,
whether the firm makes a profit or a loss,
taking into account adjustment costs.  The optimal
contract thus involves a base price which allows
the firm to break even and a premium 
for early delivery, which declines over time.

We also note that the pricing schedule
(\ref{pricing_schedule}) remains optimal from a
social welfare point of view even if there is
uncertainty about adjustment costs, which would
affect the optimal schedule in exactly the same
way for a cost minimizing firm as for a social
planner.
Given that $k$ can be assumed to be large, 
because social costs affect the entire economy, 
this strategy may lead to a high, but also quickly
declining premium. With 
such a pricing schedule there would be no 
need to specify intermediate delivery dates 
(as done in existing contracts). Firms would 
have the incentive to ramp up production 
as quickly as required by society.

%----------------------------
\subsection{Orders of magnitude 
for social costs\label{sect_orders_of_magnitude}}
%----------------------------

The orders of magnitude of the social cost of
a continuing pandemic can be estimated using the available data
on the economic cost of the pandemic so far,
which have been around 4-5 percent of GDP.
Reaching herd immunity thus allows society to avoid
costs equivalent to 4-5 percent of GDP, and even
when including value of life costs
\cite{gros2020containment}. This would
mean that the avoided economic costs per
vaccinated person would be equal to 4-5 percent
of GDP per capita. Each dose would then be worth
2-2.5 percent of GDP per capita if two doses are
needed, which would amount to 1200-1500 USD
for the US and 1000-1250 for Germany.  This
would be between 66 and 80 times the price of 15
(euro) which has been reported for a single dose
of the Pfizer/Biontech vaccine 
\cite{Vaccine_price_Biontech}. 

Another approach to determine the value of a
vaccination relies on surveys of the willingness to
pay (WTP) expressed in standard surveys used to
estimate the value of other vaccines. One study
\cite{garcia2020contingent} concludes that the
social valuation of vaccination is about 1.1
percent of the per capita gross domestic product
(GDP). This would be equivalent to about 600 USD
per dose for the US or 500 USD for Germany. These
values constitute a lower threshold as the social
value of a vaccination is likely to be substantially
larger than the private value, because vaccinated
individuals no longer transmit the disease to
others.

The estimate of the overall cost of the Covid 
pandemic presented in \cite{cutler2020covid}
suggests a similar order of magnitude, but expressed 
in total
amounts. It is estimated that the global 
total cost of the Covid pandemic is about 16 trillion 
USD (of which about one half is due to medical cost and 
the value of lives lost), which could be avoided 
through 6 billion vaccinations resulting in a social 
value of about 2600 USD per vaccination (1300 per 
dose if two are needed for immunity).

These estimates indicate that the social value for the delivery of a
vaccine today should be very high, with 1500 USD
as an upper bound, and 500 USD as a lower bound. The price a society should be willing to pay for a dose available 
immediately should be consequently between these two values.
However, the price would decline 
rapidly over time,
tending to zero towards the end of the 
delivery period (always relative to marginal costs).

The hitherto fixed prices paid by 
governments for Covid-19 vaccines 
have been made public only partially \cite{Vaccine_price_Biontech,Vaccine_prices},
but they are generally in
the region of 15-30 USD per dose.  This 
is more than one order of magnitude below 
the social value for an intermediate delivery (both for the lower and the upper bound).

%=====================================
\begin{figure}[t!]
\centerline{
\includegraphics[width=0.9\textwidth]{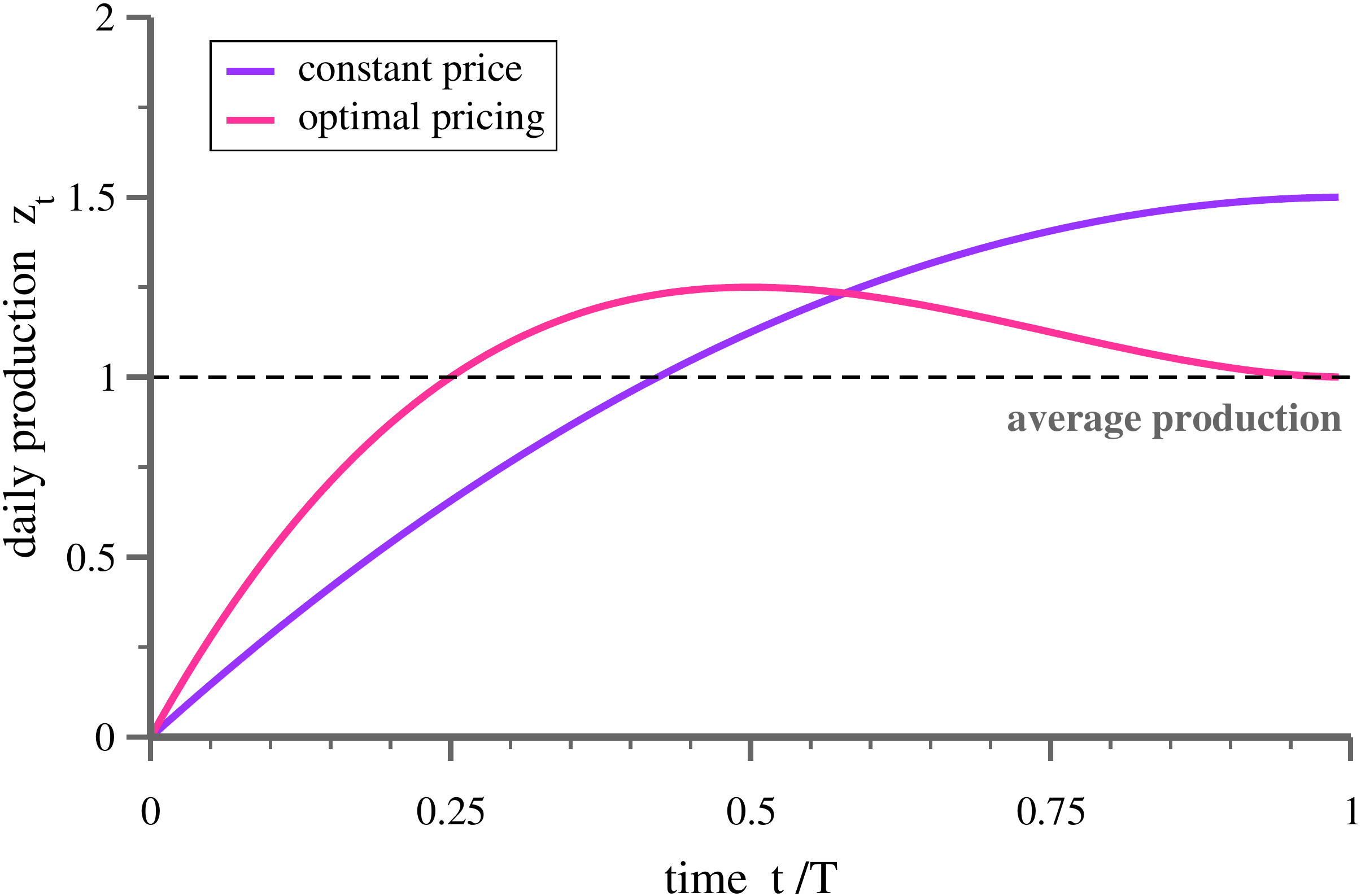}
           }
\caption{The time evolution of the production capacity
$z_t$, as for Fig.~\ref{fig_adjustmentCosts_optimal},
but with optimal pricing (\ref{pricing_schedule}),
which leads to the production timeline $z_t$ as
determined by (\ref{z1_optimal_pricing_final}) and
(\ref{gamma_delta_optimal_pricing}). Shown is the
case $\kappa=4\Delta Z/T^3$, see (\ref{z_T_nominal_production}),
for which the final production level equals the
average production $\Delta Z$. Note the temporary
overshooting. As a comparison, the result for constant
pricing, $\mu=0$ in (\ref{pricing_schedule}), is also
given.
}
\label{fig_adjustmentCosts_optimal}
\end{figure}
%=====================================

%----------------------------
\subsection{Optimal time path for capacity}
%----------------------------

The evolution of the production capacity $z_t$ optimizing
(\ref{adjustment_costs_premium}) is determined by
\begin{equation}
2a_z\ddot{z}_t = \lambda -p_0+kt,
\qquad
z_t = z_0 + \gamma t +\frac{\lambda-p_0}{4a_z}t^2 +
\frac{k}{12a_z}t^3\,,
\label{z_t_variational}
\end{equation}
where $\lambda_Z$ is again the Lagrange multiplier
enforcing the constraint that the total production
over the period $[0,T]$ is: 
\begin{equation}
Z_T = 1 = \int_0^T z_tdt =
z_0T + \frac{\gamma T^2}{2}
+\frac{(\lambda-p_0)T^3}{12a_z}
+\frac{k T^4}{48a_z}\,.
\label{Z_T_time_p}
\end{equation}
Minimizing adjustment costs implies that 
at the end of the production period it does not pay to build 
up further capacity.  This implies from (\ref{z_t_variational})
that:
\begin{equation}
0 = \dot{z}_t\Big|_{t\to T} = 
\gamma + \frac{\lambda-p_0}{2a_z}T + \frac{k}{4a_z}T^2\,.
\label{marginal_cond}
\end{equation}
The quantitative importance of a declining price schedule can be
illustrated by parameterizing $z_t$ as
\begin{equation}
z_t = z_0 + \gamma t +\delta t^2 + \kappa t^3,
\qquad\
\delta = \frac{\lambda-p_0}{4 a_z},
\qquad\
\kappa = \frac{k}{12a_z}\,.
\label{z1_optimal_pricing_final}
\end{equation}
One obtains with delivery condition 
(\ref{Z_T_time_p}) that
\begin{equation}
\gamma = \frac{2}{T^2}-\frac{2\delta T}{3}
-\frac{\kappa T^2}{2}\,,
\label{Z_T_shadow_optimal_pricing}
\end{equation}
where the initial production capacity, 
$z_0=0$, has been set to zero. We can 
now use the condition that at the end 
of the period it does not pay to build 
up further capacity:
\begin{equation}
0 = \dot{z}_t\Big|_{t\to T} = 
\gamma +2\delta T + 3\kappa T^2\,,
\label{marginal_condition}
\end{equation}
which reduces to (\ref{delta_gamma})
for $\kappa=0$. From
(\ref{Z_T_shadow_optimal_pricing}) and 
(\ref{marginal_condition}) one finds
\begin{equation}
\gamma = \frac{3}{T^2} + \frac{3\kappa T^2}{4},
\qquad\quad
\delta = -\frac{3}{2T^3} -\frac{15\kappa T}{8}\,.
\label{gamma_delta_optimal_pricing}
\end{equation}
Of interest is in particular the
choice $\kappa = 4/T^4$, which
leads to
\begin{equation}
\gamma = \frac{6}{T^2}, 
\qquad\quad
\delta = -\frac{9}{T^3}, 
\qquad\quad
z_t\big|_{t=T} = \frac{1}{T}\,.
\label{z_T_nominal_production}
\end{equation}
This result implies that the company
falls back to the average production 
at the end of the delivery period 
when $\kappa = 4/T^3$, as illustrated in
Fig.~\ref{fig_adjustmentCosts_optimal}.
At this point production is
built up at twice the speed 
resulting from constant pricing, viz
when $\kappa=0$.

%---------------------------
\subsection{Pandemic costs}
%---------------------------

The general public is directly affected in particular
by the first term of the social costs (\ref{Social_cost}),
which determines effectively when and how intensive
the vaccination campaign is. This term, the
`pandemic cost', is given by
\begin{eqnarray}
\nonumber
W_{\rm pandemic} = \int_0^T z_t t dt &=& 
\int_0^T \left(z_0 t + \gamma t^2 +
\gamma t^3 + \kappa t^4\right) dt 
\\
&=& \frac{z_0}{2} T^2 + 
\frac{\gamma}{3}  T^3 +
\frac{\delta}{4}T^4 + \frac{\kappa}{5}T^5 \,.
\label{pandemic_costs_1}
\end{eqnarray}
Using the explicitly expressions (\ref{gamma_delta_optimal_pricing}),
and $z_0=0$, the pandemic costs take the form
\begin{eqnarray}
\nonumber
W_{\rm pandemic} &= &
\left(\frac{3}{T^2} + \frac{3\kappa T^2}{4}\right)\frac{T^3}{3}
-
\left( \frac{3}{2T^3} +\frac{15\kappa T}{8} \right)\frac{T^4}{4}
+\frac{\kappa}{5}T^5 \\[0.5ex]
&=& \frac{5}{8}T - \frac{3}{160} \kappa T^5\,.
\label{pandemic_costs_2}
\end{eqnarray}
For the choice $\kappa=4/T^4$ discussed above,
see (\ref{z_T_nominal_production}), for which
the production capacity $z_t$ returns 
to the average value at the end of delivery period
$t\to T$, the pandemic costs per time, 
$W_{\rm pandemic}/T$, become $97/160$.

%===========================
\section{Conclusions}
%===========================

Our analysis starts from the observation that delays 
in the availability of vaccines are very costly for 
society. A dose delivered one quarter later is substantially
less valuable than a dose delivered today. The costs per 
unit of time remain high as the pandemic continues and 
governments are forced to implement lockdowns that depress 
the economy. However, the resulting urgency to speed up delivery 
is not recognized in the existing contracts, which 
specify mostly only a fixed quantity and an overall 
time frame, typically the entire year of 2021. 
In the absence of incentives to produce early, firms will
tend to minimize adjustment costs, viz the costs resulting
from ramping up production fast. It will then be preferable 
for firms to increase production capacity only gradually.

Our analysis shows that the lack of incentives to produce 
early does not derive from a potentially low \emph{level} of the price 
offered to companies
but on its \emph{time path}. With the existing, fixed price contracts, a dose delivered the subsequent
quarter yields the same revenue for the producer as a 
dose delivered today, but for society there is a huge difference.
The practical problem is then how to provide incentives
for early delivery. 

%One approach to reduce the discrepancy 
%between the time paths of the social value and the price
%would be to split the contract into two parts. 
%In the simplest case of a two-stage 
%contract, the first delivery batch would be higher
%priced than the second batch. For this case we present a precise relation
%between the additional costs and the resulting faster 
%production buildup. 

A better contract would have  
the price fully variable over time. 
We show that it is straightforward
to design an optimal contract, 
which aligns the time paths of the 
price with that of the social 
value of a vaccination. In this 
case linearly decreasing price 
schedules replicate the social 
optimum. 

From our perspective there is a clear policy conclusion: 
Supply contracts for vaccines should contain incentives 
for accelerated production.  Vaccines delivered early should command a higher price.

%===========================
%\section*{References}
%===========================

%%%%%%%%%%%%%%%%%%%%%%%%%%%%
%\bibliographystyle{unsrt}
%\bibliographystyle{pnas-new.bst}
%\bibliography{Corona.bib}
%%%%%%%%%%%%%%%%%%%%%%%%%%%%

\end{document}